%Paper: cond-mat/9402076
%From: mariav@hlrserv.hlrz.kfa-juelich.de (Maria Vieira )
%Date: Sat, 19 Feb 94 11:53:51 +0100

%Plain tex
\def\nms{\mathsurround=0pt}
 % greater than or approx
 % less than or approx.
\def\overapprox#1#2{\lower2pt\vbox{\baselineskip0pt\lineskip - 1pt
    \ialign{$\nms#1\hfil##\hfil$\crcr#2\crcr\approx\crcr}}}
\def\gtsim{\mathrel{\mathpalette\oversim>}} % greater than or sim.
 % less than or sim.
\def\oversim#1#2{\lower2pt\vbox{\baselineskip0pt\lineskip 1pt
    \ialign{$\nms#1\hfil##\hfil$\crcr#2\crcr\sim\crcr}}}
 % greater than or equal.
 % less than or equal.
\def\overequal#1#2{\lower2pt\vbox{\baselineskip0pt\lineskip 1pt
    \ialign{$\nms#1\hfil##\hfil$\crcr#2\crcr=\crcr}}}
\def\leftrightarrowfill{$\nms\mathord\leftarrow\mkern-6mu
       \cleaders\hbox{$\mkern-2mu\mathord- \mkern-2mu$}\hfill
       \mkern-6mu\mathord\rightarrow$}
\def\overdoublearrow#1{\vbox{\ialign{##\crcr
      \leftrightarrowfill\crcr\noalign{\kern-1pt\nointerlineskip}
        $\hfil\displaystyle{#1}\hfil$\crcr}}}
% **************************************
%
\magnification=\magstep1
\baselineskip=14pt
\centerline{\bf A GROWTH MODEL FOR DNA EVOLUTION}
\bigskip
\centerline {\bf Maria de Sousa Vieira$^*$ and Hans J. Herrmann}
\centerline {\sl Hochstleistungsrechenzentrum Supercomputing Center,
Kernforschungsanlage}
\centerline{\sl D-52425 J\"ulich }
\centerline{\sl Germany}
\bigskip
\bigskip
Peng {\sl et al.}$^1$ have reported the finding of long-range
correlations in DNA sequences, in agreement with the results of Li {\sl et
al.}$^2$
In ref. 1 the DNA sequence was analyzed by constructing a random walk
in which a pyrimidine represents a step up and a purine a step down.
{}From this walk they calculated the quantity $F(l)$ and verified that
the slope $\alpha $ of $\log F(l)$ versus $\log l$ is larger than 0.5 for
intron-containing sequences, implying the existence of
long
range correlations.
For intron-less sequences they found $\alpha =0.5$, which is
the exponent that characterizes a random sequence or a
sequence with short-range correlations.

Prabhu {\sl et al.}$^3$ and Chatzidimitriou-Dreismann {\sl et al.}$^4$ noticed
that in most cases $\log F(l)$
against $\log l$ is not a straight line. They found nonlinear
curves both for intronless sequences and those
containing introns with a local slope larger than $0.5$. According to these
results, a well-defined fractal power
exponent $\alpha $ does not usually exist for DNA sequences.

We introduce here a simple iterative model of gene evolution,
which mimics
the observed behavior of $\alpha $ in
real DNA sequences. The model incorporates the
basic features of DNA evolution, that is, sequence elongation
due to gene duplication and mutations.
In our model we start with
an intron-less sequence which consists of $N_g$ genes of equal
length. Each gene has $N_n$ nucleotides, which consists pyrimidines
and purines  randomly
distributed with the proportion of
$50 \% $ each. The process of evolution is simulated in the following
way: We choose
at random a gene of our original sequence and
in that gene we choose at random one of its nucleotides. Then,
we change (mutate) this nucleotide from a pyrimidine to a purine or
vice-versa.  A copy of the chosen gene before the mutation
is added at the end of the chain. This old gene becomes an
intron, and it is not modified anymore. The genes that
can mutate are always the first $N_g$ genes, which
are the exons in our model. We iterate this process
for $N_i$ times ($N_i >> 1$). In the
end, our chain will consist of ``head" of $N_g$ exons and a big ``tail"
of $N_i$ introns.

Next, we plot $\log F(l)$ versus $\log l$, and
as in real DNA sequences we do not find
a straight line.
The local slope $\alpha (l) $ of this curve is shown in the figure below.
The parameter values
are $N_g=10$, $N_n=3$ and $N_i=10^5,\ 5 \times 10^5, \ 10^6$, which
are represented by a solid, dotted and dashed line, respectively.
When compared with Fig. 8 of ref. 5 we observe that, for $l \gtsim N_gN_n$
(the length of our starting  sequence),
our results are in excellent agreement
with the ones obtained
from the real DNA sequence. The exponent
$\alpha $ increases monotonically from 0.5 to approximately 0.79
then decreases again to 0.5 in the limit of
very large $l$.
We have studied a large region of the parameter space and
have always found the same asymptotic behavior for $\alpha $.

In real DNA sequences, however, introns and exons are interdispersed.
For that reason, we also studied chains of the above model in
which we perform a random shuffling of exons and introns.
The only difference we noticed
for this case was that the maximum value  $\alpha $ reaches now is
about 0.62.

We also saw that if we iterate our
model by making copies of genes without any mutation
we find that  the value of $\alpha $ is constant and equal to $0.5$.
This shows that the increase of $\alpha $ is due to the mutations.

We conclude from our simple DNA growth model that mutations have the
following effect on the correlation $F(l)$ of the sequence: For
chains of intermediate size, like the ones analized in refs. 1-5,
one finds an increase in the exponent $\alpha $. This is, however,
a transient behavior since when
chains  hundred times longer ($10^6$) are analyzed, $\alpha $ is back
to $0.5$, i.e., the value of a random walk. We therefore conclude
that the interesting discovery of refs. 1 and 2 is probably the
fingerprint of the mutations that occurred during the evolution of
the DNA strand, but only a transient behaviour for not too long chains
and not a real asymptotic long-range correlation.
\bigskip
\bigskip
\item {1.} Peng {\sl et al.}, {\sl Nature} {\bf 356}, 168-170 (1992).

\item {2.} Li, W. $\& $ Kaneko, K. {\sl Europhysics Lett.} {\bf 17}, 655-660
(1992).

\item {3.} Prabhu, V. V. $\& $ Claverie, J.-M. {\sl et al.}, {\sl Nature} {\bf
359}, 782 (1992).

\item {4.} Chatzidimitriou-Dreismann, C. A. $\& $ Larhammar, D., {\sl Nature}
{\bf 361}, 212-213 (1993).

\item {5.} Buldyrev, S. {\sl et al.} {\sl Phys. Rev. E} {\bf 47}, 4514-4523
(1992).

\bigskip
\noindent{\sl $^*$Humboldt fellow}
\bigskip
\bigskip
\bigskip
\noindent {\bf Figure} - Local slope $\alpha (l)$ of $\log F(l) $ versus $\log
l$ for $N_g=10$, $N_n=3$ and
$N_i=10^5,\  5 \times 10^5,\ 10^6$, represented by a solid, dotted and dashed
line, respectively, for an ensemble average of 40 realizations.

\end